\documentclass[preprint,amsmath,amssymb,aps,prd,superscriptaddress,nofootinbib]{revtex4-1}
\usepackage{wasysym}
\usepackage{amsfonts}
\usepackage{ifsym}
\usepackage{natbib}

\def\be{\begin{equation}}
\def\ee{\end{equation}}
\def\ba{\begin{eqnarray}}
\def\ea{\end{eqnarray}}

\def\CP1{\mathbb{CP}^1}
\def\SL2C{\mathrm{SL}(2,\mathbb{C})}

\def\Z2{\mathbb{Z}_2}

\def\su2{{SU(2)}}

\def\[{\left[}
\def\]{\right]}

\def\({\left(}
\def\){\right)}
\def\[{\left[}
\def\]{\right]}
\def\lan{\langle}
\def\ran{\rangle}
\def\<{\langle}
\def\>{\rangle}

\def\i2{\frac{i}{2}}

\def\2F1{\,_2{\rm F}_1}

\def\pn{{\bf\Psi}}

\usepackage[usenames,dvipsnames]{xcolor}
\usepackage{color}
\usepackage{graphicx}
\usepackage{dcolumn}
\usepackage{bm}
\usepackage{hyperref}

\newcommand{\beq}{\begin{equation}}
\newcommand{\eeq}{\end{equation}}
\newcommand{\beqq}{\begin{equation*}}
\newcommand{\eeqq}{\end{equation*}}
\newcommand\beqa{\begin{eqnarray}}
\newcommand\eeqa{\end{eqnarray}}
\newcommand\beqaa{\begin{eqnarray*}}
\newcommand\eeqaa{\end{eqnarray*}}
\newcommand\bea{\begin{array}}
\newcommand\eea{\end{array}}

\begin{document}

%\preprint{}

\title{Einstein--Yang--Mills Scattering Amplitudes\\From Scattering Equations}


%\author{}
%\thanks{}
%\affiliation{}
\author{Freddy Cachazo}
\email{fcachazo@perimeterinstitute.ca}
\affiliation{Perimeter Institute for Theoretical Physics, Waterloo, ON N2L 2Y5, Canada}
\author{Song He}
\email{she@perimeterinstitute.ca}
\affiliation{Perimeter Institute for Theoretical Physics, Waterloo, ON N2L 2Y5, Canada}
\affiliation{School of Natural Sciences, Institute for Advanced Study, Princeton, NJ 08540, USA}
\author{Ellis Ye Yuan}
\email{yyuan@perimeterinstitute.ca}
\affiliation{Perimeter Institute for Theoretical Physics, Waterloo, ON N2L 2Y5, Canada}
\affiliation{Department of Physics \& Astronomy, University of Waterloo, Waterloo, ON N2L 3G1, Canada}


\date{\today}

\begin{abstract}

We present the building blocks that can be combined to produce tree-level S-matrix elements of a variety of theories with various spins mixed in arbitrary dimensions. The new formulas for the scattering of $n$ massless particles are given by integrals over the positions of $n$ points on a sphere restricted to satisfy the scattering equations. As applications, we obtain all single-trace amplitudes in Einstein--Yang--Mills (EYM) theory, and generalizations to include scalars. Also in EYM but extended by a B-field and a dilaton, we present all double-trace gluon amplitudes. The building blocks are made of Pfaffians and Parke--Taylor-like factors of subsets of particle labels.

\end{abstract}

%\pacs{}
%\keywords{}

\maketitle

\section{Introduction}

The complete tree-level S-matrices of Einstein gravity, pure Yang--Mills and cubic massless scalars in arbitrary dimensions admit compact representations as integrals over the moduli space of a punctured sphere \cite{Cachazo:2013hca,Cachazo:2013iea}. The key ingredient in the construction is the scattering equations
\be
\sum_{b\neq a} \frac{s_{ab}}{\sigma_{a}-\sigma_{b}} = 0 \quad {\rm for} \quad a\in \{ 1,2,\ldots , n\}
\label{scatt}\ee
where $s_{ab}:=(k_a+k_b)^2=2\,k_a\cdot k_b$, and $\sigma_c$ is the position of the $c^{\rm th}$ puncture. These equations have made an appearance at various times in the literature in a variety of different contexts~\cite{Fairlie:1972, *Roberts:1972, *Fairlie:2008dg, Gross:1987ar, Witten:2004cp, Caputa:2011zk, *Caputa:2012pi, Makeenko:2011dm,Cachazo:2012uq}. The scattering equations generically have $(n-3)!$ solutions and admit a polynomial form which simplifies their solution \cite{Dolan:2014ega}.

The S-matrices are given by
\be  {\cal M}_n = \int \frac{d\,^n\sigma}{\textrm{vol}\,\SL2C} \prod_a {}'\delta(\sum_{b\neq a} \frac{s_{ab}}{\sigma_{a}-\sigma_{b}})~{\cal I}_n(\{k,\epsilon,
\sigma\}),\label{generalformula}
\ee
where ${\cal I}_n(\{k,\epsilon,
\sigma\})$ is an integrand that depends on the theory and carries all the information about wave functions for the external particles, i.e., polarization vectors and tensors for gluons and gravitons respectively. In this formula $\prod_a {}'$ refers to the fact that three delta functions must be removed in a way explained in \cite{Cachazo:2013hca,Cachazo:2013iea} and reviewed in Section \ref{examples}.

In the original construction, known as CHY formulas, two building blocks were identified~\cite{Cachazo:2013hca}:
\be\label{bint}
C(1,2,\ldots, n) = \frac{1}{\sigma_{12}\sigma_{23}\cdots \sigma_{n1}}, \quad
E(\{k,\epsilon ,\sigma\} ) = {\rm Pf}'\Psi(\{k,\epsilon ,\sigma\} )
\ee
where $\Psi(\{k,\epsilon ,\sigma\} )$ is a $2n\times 2n$ matrix whose structure we review in Section \ref{section2} and $\sigma_{ab}$ is a shorthand notation for $\sigma_a-\sigma_b$. Combining any two of the blocks (where repeating one is allowed) one produces amplitudes of physical theories,
\ba\label{build}
\nonumber {\cal I}^{\rm gravity}_n &=& E(\{k,\epsilon ,\sigma\} )^2, \\
\nonumber {\cal I}^{\rm Yang-Mills}_n &=& C(1,2,\ldots, n)E(\{k,\epsilon ,\sigma\} ), \\
{\cal I}^{\rm scalar}_n &=& C(1,2,\ldots, n)^2.
\ea
For more details we refer the reader to \cite{Cachazo:2013iea} where relations to Kawai--Lewellen--Tye (KLT) \cite{Kawai:1985xq} and Bern--Carrasco--Johansson (BCJ) double copy constructions \cite{Bern:2008qj} were also studied.

Of course, complete Yang--Mills amplitudes with group $U(N)$ are obtained by using
\be\label{color}
{\cal C}_n = \sum_{\omega \in S_n/{\mathbb{Z}_n}}\frac{{\rm Tr}(T^{a_{\omega(1)}}T^{a_{\omega(2)}}\cdots T^{a_{\omega(n)}})}{\sigma_{\omega(1)\omega(2)}\sigma_{\omega(2)\omega(3)}\cdots \sigma_{\omega(n)\omega(1)}}
\ee
instead of a single $C$ which gives rise to what is known as a partial amplitude.

Several generalizations of the original formulas have been proposed in the literature. These include the extension to massive scalar particles with cubic interactions \cite{Dolan:2014ega} and to amplitudes with two massive scalars and the rest gluons or gravitons \cite{Naculich:2014naa}.

The striking similarity of all these constructions to string theory ones led Mason and Skinner to the construction of a series of ambitwistor string theories whose correlation functions are localized on the scattering equations \cite{Mason:2013sva}. A type II version leads to a consistent theory with correlators that compute scattering amplitudes of gravitons in a form that matches exactly the CHY formula. In a follow-up work \cite{Adamo:2013tsa} three and four point amplitudes mixing gravitini and gravitons were also presented in compact form along with a generalization to one loop. In \cite{Mason:2013sva}, a purely bosonic theory and a heterotic version were, unfortunately, found not to correspond to any known gravity theory. However, the purely gluonic single-trace amplitudes in the heterotic theory give rise to correct CHY formulas. For other interesting developments related to ambitwistor formulations specializing to four dimensions, see \cite{Bandos:2014lja, Geyer:2014fka}. While these constructions are based on the analog of the RNS formalism for strings, a version based on pure spinors was constructed by Berkovits \cite{Berkovits:2013xba} and further studied in \cite{Gomez:2013wza}.

In this paper we continue the search for more general formulations based on the scattering equations which allow the description of general physical theories. The key observation is that the building blocks \eqref{build} can be replaced by products of smaller ones. The new building blocks are nothing but copies of those in \eqref{build} but defined for subsets of particles.

The main result of this work is a formula for the single trace amplitude of $r$ gluons and $s$ gravitons ($n=r+s$) in Einstein--Yang--Mills (EYM) theory:
\be\label{EYM}
{\cal I}^{\rm EYM}_{r,s} ={\cal C}_r~E(\epsilon_{r+1},\epsilon_{r+2}, \ldots ,\epsilon_n)~E(\epsilon_1,\epsilon_2,\ldots ,\epsilon_n)
\ee
where ${\cal C}_r$ is given by \eqref{color} but only for particles $\{1,2,\ldots ,r\}$, and the precise form of $E(\epsilon_{r{+}1},\epsilon_{r{+}2}, \ldots,\epsilon_n)$ will be given in the next section.

We also propose a formula for all double--trace gluon amplitudes in EYM extended by a B-field and a dilaton. The formula for two traces with orderings $(1,2,\ldots ,p)$ and $(p+1,p+2,\ldots ,n)$ is given by
\be\label{EYMDT}
{\cal I}^{{\rm EYM}(2)}_{1,2,\dots ,p ; p+1,\ldots n} = C(1,2,\ldots ,p)~s_{12\ldots p}~C(p+1,p+2,\ldots ,n)~E(\epsilon_1,\epsilon_2,\ldots ,\epsilon_n)
\ee
where $s_{12\ldots p}=(k_1+k_2+\cdots +k_p)^2$ and the superscript $(2)$ indicates the number of traces.
  
We end the illustrations of the use of the building blocks with formulas for a scalar minimally coupled to Yang--Mills (YMS). Finally, using the KLT \cite{Kawai:1985xq} construction we show how to obtain the general formula for Einstein--Yang--Mills--scalar theory (EYMS); in particular \eqref{EYM} can be derived by combining the formula for YMS with that for gluon amplitudes in pure Yang--Mills. We end with a section on discussions and future directions.

\section{Building Blocks}\label{section2}

Let us define a set of scattering data for $n$ massless particles as a collection of momentum vectors $\{ k_1,k_2,\ldots ,k_n\}$ satisfying $k_a^2=0$ and momentum conservation. Also in the data is a set of polarization vectors $\{ \epsilon_1,\epsilon_2,\ldots ,\epsilon_n\}$.

The first set of building blocks is obtained by considering a subset %${\cal S}$ of labels of $\{ 1,2,\ldots ,n\}$ with $r$ elements and $r\geq n$ to define
${\cal S}=\{i_1,i_2,\ldots,i_r\}\subseteq\{ 1,2,\ldots ,n\}$ to define
\be
{\cal C}_{\cal S} = \sum_{\omega \in S_r/{\mathbb{Z}_r}}\frac{{\rm Tr}(T^{a_{\omega(i_1)}}T^{a_{\omega(i_2)}}\cdots T^{a_{\omega(i_r)}})}{\sigma_{\omega(i_1)\omega(i_2)}\sigma_{\omega(i_2)\omega(i_3)}\cdots \sigma_{\omega(i_r)\omega(i_1)}}
\ee\label{defC}
where $2\leq r\leq n$ is the number of elements in ${\cal S}$, and for $r=0$, $\mathcal{C}_{\emptyset}:=1$. It is useful to define the partial amplitude version, which is nothing but the formula \eqref{bint} applied to the set ${\cal S}$,
\be
C(i_1,i_2,\ldots ,i_r) = \frac{1}{\sigma_{i_1i_2}\sigma_{i_2i_3}\cdots \sigma_{i_ri_1}}.
\ee

The second set of building blocks is also defined for a given subset ${\cal S}$. In this case one defines an anti-symmetric $2r\times 2r$ matrix, $\Psi_{\cal S}$, as follows
\be\label{Psi}
\Psi_{\cal S} = \left(
         \begin{array}{cc}
           A &  -C^{\rm T} \\
           C & B \\
         \end{array}
       \right)
\ee
where $A$, $B$ and $C$ are $r\times r$ matrices. The first two matrices have components
\be
A_{ab} = \begin{cases} \displaystyle \frac{s_{ab}}{\sigma_{a}-\sigma_{b}} & a\neq b,\\
\displaystyle \quad ~~ 0 & a=b,\end{cases} \qquad B_{ab} = \begin{cases} \displaystyle \frac{2\,\epsilon_a\cdot\epsilon_b}{\sigma_{a}-\sigma_{b}} & a\neq b,\\
\displaystyle \quad ~~ 0 & a=b,\end{cases}
\label{ABmatrix}
\ee
while the third is given by
\be
C_{ab} = \begin{cases} \displaystyle \frac{2\,\epsilon_a \cdot k_b}{\sigma_{a}-\sigma_{b}} &\quad a\neq b,\\
\displaystyle -\sum_{j=1;j\neq a}^n\frac{2\,\epsilon_a \cdot k_j}{\sigma_{a}-\sigma_{j}} &\quad a=b.\end{cases}
\ee
In these matrices we choose to label the entries not by the standard $\{ 1,2,\ldots ,r\}$ labels but by the labels in the set ${\cal S} = \{ i_1,i_2,\ldots ,i_r \}$. When $1\leq r\leq n-2$ the matrix $\Psi_{\cal S}$ is non-degenerate and we define
\be
E(\epsilon_{i_1},\epsilon_{i_2},\ldots ,\epsilon_{i_r}) = {\rm Pf}\Psi_{\cal S}.
\ee
In the case $r=n-1$ we define $E=0$ and when $r=n$ we use the definition given in \cite{Cachazo:2013hca} for the computation of pure Yang--Mills amplitudes. For the reader's convenience we recall the definition here. Let $\pn$ denote the matrix $\Psi_{\cal S}$ for the set ${\cal S}=\{ 1,2,\ldots ,n\}$ and $\pn_{ij}^{ij}$ the matrix obtained from $\pn$ by deleting rows $\{i,j\}$ and columns $\{i,j\}$. Here $\{i,j\}$ are chosen from the first $n$ rows and the first $n$ columns. In~\cite{Cachazo:2013hca} it was noted that $\Psi$ has corank 2 and therefore its Pfaffian vanishes. The correct definition of $E$ is the so-called reduced Pfaffian
\be\label{redPf}
E(k,\epsilon,\sigma ) ={\rm Pf}'\pn := \frac{(-1)^{i+j}}{\sigma_{ij}}{\rm Pf}\pn_{ij}^{ij}.
\ee
It was also proven in \cite{Cachazo:2013hca} that \eqref{redPf} is independent of the choice of $\{i,j\}$. This is the object that appears in the formulas, \eqref{build}, for pure Yang--Mills and pure gravity presented in the Introduction.

\section{Einstein--Yang--Mills Scattering Amplitudes}

In this section we focus on the scattering amplitudes of Yang--Mills theory with gauge group $U(N)$ minimally coupled to Einstein gravity (EYM). The scattering amplitudes can be classified by the total number of particles $n$, the number of gluons $r$ and the number of gravitons $s$. Clearly $r+s=n$. A further classification is possible by expressing the structure constants of $U(N)$ in terms of traces of generators $T^a$ and performing a color decomposition. This procedure leads to amplitudes with a single trace, double traces, etc. In this paper we study all single-trace amplitudes, leaving multiple-trace ones for future work. Assuming, without loss of generality, that the gluons are particles $\{1,2,\ldots ,r\}$ and the gravitons the rest, one has
\be\label{ourF}
{\cal I}^{\rm EYM}_{r,s} ={\cal C}_r~E(\epsilon_{r+1},\epsilon_{r+2}\ldots ,\epsilon_n)~E(\epsilon_1,\epsilon_2,\ldots ,\epsilon_n).
\ee
Let us give some examples which also illustrate the definitions of the building blocks.

\subsection{Examples}\label{examples}

Consider the case when $r=n-1$ and $s =1$, i.e., a single graviton and the rest gluons. In this case $E(\epsilon_n)$ is constructed from a $2\times 2$ matrix. The matrix $\Psi_{\cal S}$ is simply
\be\label{PsiEYM}
\Psi_{\cal S} = \left(
         \begin{array}{cc}
           0 &  -C_{nn} \\
           C_{nn} & 0 \\
         \end{array}
       \right),
\ee
where
\be\label{cnn}
C_{nn} =- \sum_{a=1}^{n-1}\frac{2\,\epsilon_n\cdot k_a}{\sigma_{na}}.
\ee
This means that $E(\epsilon_n) ={\rm Pf}\Psi_{\cal S} = C_{nn}$. Now we can explicitly write a partial amplitude with the standard ordering for the gluons $\{1,2,\ldots ,n-1\}$ as
\be\label{exp1}
A^{\rm EYM}_{n-1,1} = \int \frac{d\,^n\sigma}{\textrm{vol}\,\SL2C} \prod_a {}'\delta(\sum_{b\neq a} \frac{s_{ab}}{\sigma_{a b}})\frac{C_{nn}~{\rm Pf}'\pn}{\sigma_{12}\sigma_{23}\cdots \sigma_{n-1,1}}.
\ee
In order to use this formula in explicit computations it is important to explain the meaning of various structures. The reason one has to divide by the volume of $\SL2C$ is that the integrand is invariant under such transformations. Gauge-fixing any three variables, say $\sigma_x,\sigma_y$ and $\sigma_z$ one can substitute
\be
\int \frac{d\,^n\sigma}{\textrm{vol}\,\SL2C}   \longrightarrow \int\!\! \prod_{a=1; a\neq x,y,z}^n \!\!\!\! d\,\sigma_a ~ \sigma_{xy}\sigma_{yz}\sigma_{zx}.
\ee
The last ingredient is the definition of $\prod_a {}'$. The reason a special definition is needed is that if we denote
\be
Q_a = \sum_{b=1;b\neq a}^n \frac{s_{ab}}{\sigma_{a b}}
\ee
then it is easy to show that $\sum_a \sigma_a^l~Q_a = 0$ for $l\in \{ 0,1,2\}$. This means that three of the $n$ delta functions are redundant. It turns out that removing any three $Q$'s, say, $Q_i,Q_j$ and $Q_k$ one can defined an object independent of the choice as follows,
\be
\prod_a {}'\delta(\sum_{b\neq a} \frac{s_{ab}}{\sigma_{a b}}) = \sigma_{ij}\sigma_{jk}\sigma_{ki}\prod_{a=1;a\neq i,j,k}^n \delta(\sum_{b\neq a} \frac{s_{ab}}{\sigma_{a b}}).
\ee
Combining all these ingredients one finds that \eqref{exp1} can be evaluated by summing over the $(n-3)!$ solutions to the scattering equations. The complete amplitude in the single-trace sector is obtained by dressing the above result with the trace of the generators in the canonical ordering and summing over $(n-1)!$ permutations modulo cyclic ones.

Consider now the case with two gravitons, i.e., $r=n-2$ and $s=2$. The relevant matrix for ${\cal S}=\{n-1,n\}$ is given by
\be\label{PsiEYM2}
\Psi_{\cal S} = \left(
         \begin{array}{cccc}
           0 & \frac{s_{n-1,n}}{\sigma_{n-1,n}} &  -C_{n-1,n-1} & -\frac{2\,\epsilon_n\cdot k_{n-1}}{\sigma_{n,n-1}} \\
           \frac{s_{n,n-1}}{\sigma_{n,n-1}} & 0 &  -\frac{2\,\epsilon_{n-1}\cdot k_{n}}{\sigma_{n-1,n}} &      -C_{nn} \\ C_{n-1,n-1} & \frac{2\,\epsilon_{n-1}\cdot k_{n}}{\sigma_{n-1,n}} & 0 & \frac{2\,\epsilon_{n-1}\cdot \epsilon_{n}}{\sigma_{n-1,n}} \\ \frac{2\,\epsilon_n\cdot k_{n-1}}{\sigma_{n,n-1}} & C_{nn} & \frac{2\epsilon_{n}\cdot \epsilon_{n-1}}{\sigma_{n,n-1}} & 0
         \end{array}
       \right),
\ee
and the amplitude reads
\be
A^{\rm EYM}_{n-2,2} = \int \frac{d\,^n\sigma}{\textrm{vol}\,\SL2C} \prod_a {}'\delta(\sum_{b\neq a} \frac{s_{ab}}{\sigma_{a b}})\frac{{\rm Pf}\Psi_{\cal S}~{\rm Pf}'\pn}{\sigma_{12}\sigma_{23}\cdots \sigma_{n-2,1}}.
\ee

\section{Consistency Checks}

In this section we present consistency checks for our formula \eqref{color}. %Some trivial but important properties are the following. Our proposal is permutation invariant in the graviton labels and cyclic invariant in the gluon labels.
First of all, it is straightforward to observe that this formula is permutation invariant in the graviton labels and cyclic invariant in the gluon labels, as expected. It is also multilinear in the polarization vectors of the gluons, as well as in the polarization tensors of the gravitons defined as $\epsilon_a^{\mu\nu} = \epsilon^\mu_a\epsilon_a^\nu$.

\subsection{Gauge Invariance}

A very crucial property of physical amplitudes of gluons and gravitons is gauge invariance, in order that the resulting object transforms correctly under Poincar\'{e} transformations. In the case of gluons this requires that if any polarization vector $\epsilon_a$, with $a\in \{1,2,\ldots ,r\}$, is replaced by its corresponding momentum $k_a$, then the amplitude must vanish. This is obvious since the matrix $\pn$ develops an additional null vector (two rows and hence two columns become identical) and therefore ${\rm Pf}'\pn \varpropto {\rm Pf}\pn_{ij}^{ij} = 0$. The case of gravitons is more interesting. Here we take $a\in \{ r+1,\ldots ,n\}$ and write
\ba
\nonumber E(\epsilon_{r+1},\epsilon_{r+2}\ldots ,\epsilon_n) & = & {\rm Pf}\Psi_{\cal S} =  \epsilon_a^\mu (T_{\cal S})_\mu, \\
E(\epsilon_1,\epsilon_2,\ldots ,\epsilon_n) & = & {\rm Pf}'\pn = \epsilon_a^\nu T_\nu.
\ea
The product then can be written as
\be
\epsilon^{\mu\nu}_a (T_{\cal S})_\mu T_\nu.
\ee
Gauge invariance is the statement that the amplitude must vanish if $\epsilon^{\mu\nu}_a$ is replaced by $\Lambda^\mu k_a^\nu + k_a^\mu \Lambda^\nu$. The vanishing of the amplitude is also clear in this case since $T_\nu k_a^\nu = 0$ as in the gluon case and $(T_{\cal S})_\mu k_a^\mu = 0$ for the same reason, i.e., the matrix $\Psi_{\cal S}$ develops a null vector that it did not have before.

\subsection{Soft Limits}

There are two types of soft limits: either a graviton or a gluon becomes soft. Let us start with a graviton soft limit. In our
notation, particle $n$ is always a graviton if there is at least one in the amplitude. Since our formula is permutation invariant in the graviton labels, the $n^{\rm th}$ graviton is not a special choice. Consider the limit when $k_n\to 0$. Just as in the case of pure Yang--Mills or pure gravity, the reduced Pfaffian of the matrix $\pn_n$ becomes
\be
{\rm Pf}'\pn_n = C_{nn}{\rm Pf}'\pn_{n-1}
\ee
Here we have added the subscripts $n$ and $n-1$ to indicate that the matrices correspond to amplitudes with $n$ or $n-1$ particles respectively. $C_{nn}$ is defined in \eqref{cnn}.

Very nicely, the same is true for the Pfaffian of $\Psi_{\cal S}$ where ${\cal S} =\{ r+1,\ldots ,n-1,n\}$,
\be
{\rm Pf}\Psi_{\cal S} = C_{nn}{\rm Pf}\Psi_{\hat {\cal S}}
\ee
and $\hat {\cal S} =\{ r+1,\ldots ,n-1\} $.

Returning to the complete formula for the amplitude one has
\be
A^{\rm EYM}_{r,s} = \int \frac{d\,^{n-1}\sigma}{\textrm{vol}\,\SL2C} \prod_{a=1}^{n-1} {}'\delta(\sum_{b=1,b\neq a}^{n-1} \frac{s_{ab}}{\sigma_{a b}})\frac{{\rm Pf}\Psi_{\hat {\cal S}}{\rm Pf}'\pn_{n-1}}{\sigma_{12}\sigma_{23}\cdots \sigma_{r 1}}\,\int d\sigma_n \delta (\sum_{b=1}^{n-1}\frac{s_{na}}{\sigma_{na}})C_{nn}^2.
\ee
A simple residue theorem argument shows that
\be
\int d\sigma_n \delta (\sum_{b=1}^{n-1}\frac{s_{na}}{\sigma_{na}})C_{nn}^2 = \sum_{a=1}^{n-1}\frac{(\epsilon_n\cdot k_a)^2}{k_n\cdot k_a},
\ee
which factors out of the other integrals and gives the expected result from Weinberg's soft theorem~\cite{Weinberg:1964ew,Weinberg:1965nx}
\be
A^{\rm EYM}_{r,s} \longrightarrow \left( \sum_{a=1}^{n-1}\frac{(\epsilon_n\cdot k_a)^2}{k_n\cdot k_a} \right)~ A^{\rm EYM}_{r,s-1}.
\ee

Finally we consider a soft gluon. Let the soft gluon be particle $r$. The behavior of ${\rm Pf}'\pn_n$ is the same as before, i.e.,
\be
{\rm Pf}'\pn_n = C_{rr}{\rm Pf}'\pn_{n-1},
\ee
while ${\rm Pf}\Psi_{\cal S}$ does not produce any additional factor. Moreover all dependence on particle $r$ completely drops out of ${\rm Pf}\Psi_{\cal S}$ in the limit $k_r\to 0$.

Next, we write the Parke--Taylor-like factor as
\be
\frac{1}{\sigma_{12}\sigma_{23}\cdots \sigma_{r,1}} = \frac{1}{\sigma_{12}\sigma_{23}\cdots \sigma_{r-1,1}}\frac{\sigma_{r-1,1}}{\sigma_{r-1,r}\sigma_{r,1}}.
\ee
Combining all the pieces together we are faced with the contour integral
\be
\int d\sigma_n \delta (\sum_{b=1}^{n-1}\frac{s_{na}}{\sigma_{na}})C_{rr}\frac{\sigma_{r-1,1}}{\sigma_{r-1,r}\sigma_{r,1}}.
\ee
Once again a residue theorem gives rise to the correct soft factor.

\subsection{Explicit Comparison with Known Amplitudes}\label{explicitchecks}

In addition to the above, further consistency checks are provided by comparisons between results from our formula and amplitudes that either exist in the literature or can be easily computed using Feynman diagrams.

Luckily, all amplitudes with one graviton and the rest gluons have been studied in the literature~\cite{Chen:2010ct, Stieberger:2014cea}, which can be derived from the so-called ``disk relations" in string theory~\cite{Chen:2009tr, Stieberger:2009hq, Chen:2010sr}. In the field-theory limit, these relations express such amplitudes as linear combinations of pure Yang--Mills amplitudes where the graviton is replaced by two collinear gluons \cite{Stieberger:2014cea}. The linear combinations only involve partial Yang--Mills amplitudes where the new gluons, i.e., $n$ and $n{+}1$, are never adjacent, which means that these partial amplitudes are finite in the collinear limit. %Some partial results in this direction were already obtained earlier by Stieberger in \cite{Stieberger:2009hq} as well as some results on the case of two gravitons.

In the case of one graviton, one can analytically prove that, the formula explicitly given in~\cite{Stieberger:2014cea} is equivalent to our formula \eqref{exp1}. In order to see this note that when taking the collinear limit $k_n \cdot k_{n{+}1}\to 0$ of the Yang--Mills formula, we have $\sigma_{n{+}1}\to \sigma_n$, and the reduced Pfaffian becomes
\be
{\rm Pf}' \pn_{n{+}1} \to C_{n n}\,{\rm Pf}~ \pn_n\,,
\ee
where we take $\epsilon_{n{+}1}=\epsilon_n$. Furthermore, by applying the linear combination given in~\cite{Stieberger:2014cea} to the Parke--Taylor-like objects, we find the expected $C(1,\ldots,n{-}1)$, and a pre-factor that cancels exactly the Jacobian from integrating out $\sigma_{n{+}1}$, thus we arrive at \eqref{exp1}.  When restricted to four dimensions, we also compared the two formulas numerically for all helicity sectors up to $n=8$ points.

We have also checked our formula for multiple-graviton cases. In four dimensions, closed formulas for $A^{\rm EYM}_{r,s}$ are available when we restrict to the MHV sector with two negative-helicity particles being either two gluons or a gluon and a graviton~\cite{Bern:1999bx}. %The formulas were obtained by %KLT relations from pure Yang-Mills amplitude and scalar-YM mixed amplitude, and the latter can be %determined by Feynman diagrams.
Comparison with this formula was performed numerically up to $n=7$ for all possible pairs $(r,s)$.

All checks that we have performed show complete agreement. In fact, it should be possible to directly show that our formula satisfies the factorization properties of EYM amplitudes, but for simplicity, we restrict the study of factorizations to the case of the Yang--Mills--scalar theory (see section \ref{fac}).

\section{Double-trace Gluon Amplitudes in Einstein--Yang--Mills}

In this section we consider the first generalizations of our formula beyond single-trace amplitudes considered above. In EYM, gluons belonging to different traces can interact via the exchange of gravitons. The theory we consider here is slightly more general than the pure EYM, where we can also exchange dilatons and B-fields all coupled in the standard way. In a slight abuse of terminology, we will still refer to this theory as Einstein--Yang--Mills (EYM).

We focus on the case when external legs are all gluons and distributed into two traces. Without loss of generality, the sets of particles in the two traces can be chosen as $\{1,\ldots, p\}$ and $\{p{+}1, \ldots, n\}$. We propose a formula for the double-trace gluon amplitudes, as \eqref{generalformula} with the integrand given by,
\be\label{doubletrace}
{\cal I}^{{\rm EYM}(2)}_{1,2,\ldots ,p;,p+1,\ldots ,n}={\cal C}_{\{1,\ldots, p\}}\, s_{12\ldots p}\,{\cal C}_{\{p{+}1, \ldots, n\}}\,E(\{k,\epsilon ,\sigma\} )\,,
\ee
where the definition of ${\cal C}_{\cal S}$ for a set ${\cal S}$ is given in \eqref{defC}, and $s_{12\ldots p}=(k_1+k_2+\cdots +k_p)^2$. Equivalently we can color-decompose the amplitude as
\be
\mathcal{A}^{{\rm EYM}(2)}(\{1,\ldots,p\} ;\{p{+}1,\ldots,n\})= \!\!\! \!\sum_{\substack{\omega \in S_{p{-}1} \\
\tau\in S_{n{-}p{-}1}}}
\!\!\!{\rm Tr}(T^{a_{\omega(1)}}\cdots T^{a_{\omega(p)}})\,{\rm Tr}(T^{a_{\tau(p{+}1)}}\cdots T^{a_{\tau(n)}})\, A^{{\rm EYM}(2)}(\omega; \tau)\,,\nonumber
\ee
where $A^{{\rm EYM}(2)}(\omega; \tau)$ is a shorthand notation for the double-trace partial amplitudes with orderings $\omega$ and $\tau$, $A^{{\rm EYM}(2)}(\omega(1),\ldots, \omega(p); \tau(p{+}1), \ldots, \tau(n))$.
The integrand for $A^{{\rm EYM}(2)}(\omega; \tau)$ contains two corresponding Parke-Taylor-like factor,
\be\label{doubletracepartial}
{\cal I}^{{\rm EYM}(2)}(\omega ; \tau) = \frac{1}{\sigma_{\omega(1),\omega(2)} \cdots \sigma_{\omega(p), \omega(1)}}~s_{12\ldots p}~\frac{1}{ \sigma_{\tau(p{+}1),\tau(p{+}2)} \cdots \sigma_{\tau(n),\tau(p{+}1)}}E(\{k,\epsilon ,\sigma\} )\,.
\ee

%
%\be\label{doubletracepartial}
%{\cal I}^{{\rm EYM}(2)}(\omega ; \tau) = \frac{ s_{1,\ldots,p} \, E(\{k,\epsilon ,\sigma\} )}{\sigma_{\omega(1),\omega(2)} \cdots \sigma_{\omega(p), \omega(1)} \sigma_{\tau(p{+}1),\tau(p{+}2)} \cdots \sigma_{\tau(n),\tau(p{+}1)}}\,.
%\ee

The formulas, \eqref{doubletrace} and \eqref{doubletracepartial}, have the correct symmetry in both traces, and their multilineariry in polarizations and gauge invariance are identical to the single-trace case. Compared to the single-trace gluon amplitude case, an additional Mandelstam variable is expected purely from dimensional analysis: with the exchange of one graviton (or dilation/B field), we need an additional factor with mass-squared dimension for the double-trace amplitude. The Mandelstam variable $s_{12\ldots p}$ is the simplest option with the expected symmetry between the two sets of particles. More importantly, it can be checked explicitly that near the factorization channel where $s_{12\ldots p}\to 0$, the explicit factor in the numerator turns a double pole into a single pole as expected.
The argument for checking the correct soft gluon limits is the same as above, except that the two contributions in the soft factor, for a partial amplitude, come from the two neighboring legs within the trace of the soft leg. 

Explicit comparisons with known amplitudes were also performed. Here we restrict to four dimensions. The simplest case is that of MHV amplitudes, which can be obtained from BCFW recursion relations,
\be
A^{{\rm EYM}(2)}(\omega; \tau | i^-,j^-)=\frac{ s_{12\ldots p}\,\lan i j\ran^4}{\lan \omega(1),\omega(2)\ran \cdots \lan \omega(p), \omega(1)\ran\, \lan \tau(p{+}1),\tau(p{+}2)\ran \cdots \lan \tau(n),\tau(p{+}1)\ran},
\ee
where we have negative helicity gluons $i,j$. Similar to the single-trace case, one can analytically show that \eqref{doubletracepartial} reduces exactly to this expression. In addition, it is also straightforward to use BCFW to obtain NMHV six-gluon amplitudes for $p=2,3$. For example, the two inequivalent helicity amplitudes for $p=3$ are $A_{\rm NMHV}^{{\rm EYM}(2)}(1^-,2^-, 3^-; 4^+, 5^+, 6^+)$ and  $A^{{\rm EYM}(2)}_{\rm NMHV}(1^-,2^-, 3^+; 4^-, 5^+, 6^+)$. We have checked numerically that in all cases our formula gives the correct result.

\section{Yang--Mills--Scalar Amplitudes and Generalizations}

In this section we consider the inclusion of scalars, in the adjoint of the product of two color groups, $U(N)\times U(\tilde N)$, which are minimally coupled to Einstein gravity, and to Yang--Mills with gauge group $U(N)$. Again we restrict to the single-trace case, and we first consider the case with scalars and gluons, which will be referred to as Yang--Mills--Scalar theory (YMS).

The interaction vertices of the theory can be found in~\cite{Bern:1999bx}, where the scalars are coupled to gluons with the same coupling constant as the Yang--Mills coupling. Given this, one can construct EYM amplitudes by applying KLT relations with YMS and pure gluon partial amplitudes~\cite{Bern:1999bx}. Note that partial amplitudes in YMS are defined with respect to one copy of the gauge groups, e.g., $U(N)$, and the scalars still carry the color indices of $U(\tilde{N})$.

Here we first propose the formula for amplitudes in YMS, which can be used to give the most general formula via KLT relations. For $q$ scalars and $r$ gluons with $q+r=n$, the integrand is
\be~\label{scaYM}
\mathcal{I}^{\rm{YMS}}_{q,r}=\mathcal{C}_n^{U(N)}\, \mathcal{C}_q^{U(\tilde N)}\, E(\epsilon_{q{+}1},\ldots, \epsilon_n),
\ee
where we have indicated the gauge groups: $U(N)$ for all particles, and $U(\tilde N)$ for scalars only. Properties such as gauge invariance and soft limits can be checked as above, and we point out that for the special case $q=2$, the formula coincides with the massless limit of the one for two scalars and the rest gluons as presented in \cite{Naculich:2014naa}\footnote{There is a subtlety that in \cite{Naculich:2014naa} the scalar is in the fundamental representation, but this only leads to a difference in the color structure.}. In addition, in four dimensions we have checked our formula numerically up to $n=6$,  against results obtained from Feynman diagrams or BCFW recursions, for all possible pairs $(q,r)$ and all color orderings.

More interestingly, with small number of gluons, it is straightforward to write down the explicit result from \eqref{scaYM} in any dimensions. For example, if we focus on the canonical color ordering, and $r=2$ with non-adjacent gluons $i$, $j$, we have
\be
{\rm Pf} \Psi_{\mathcal{S}}=\frac{4\,\epsilon_i \cdot\epsilon_j\, k_i \cdot k_j}{\sigma_{i j}^2}-\frac{4\,\epsilon_i\cdot k_j\, \epsilon_j\cdot k_i}{\sigma_{i j}^2}-C_{ii}\,C_{jj}.
\ee
By properly applying partial fraction relations on terms in $C_{ii}$ and $C_{jj}$, we find an expression where the coefficient of any contraction corresponds to a specific double-partial scalar amplitude $m(\alpha|\beta)$, which can be read off from the integrand directly (see \cite{Cachazo:2013iea}). The complete analytic result valid in any dimensions can thus be obtained, which can be compared with the result computed from Feynman diagrams directly. For up to $n=6$ with $r=2$ non-adjacent gluons, we find complete agreement and we expect it continue to hold for higher points. Now we turn to the study of factorizations of our formula.

\subsection{Factorizations of Yang--Mills--Scalar formula}\label{fac}

%Unitarity in YMS is much richer as compared to, e.g., pure Yang-Mills, since in general one expects to see both scalar and gluon propagating upon a given factorization channel, which imposes a strong consistency constraint on the amplitude formula.
%
%We again focus only on single-trace amplitudes. From the Feynman diagram point of view, since the $U(N)$ part of the scalar color group is not shared with the gluons, the single-trace condition requires that in any amplitude all the scalar lines should form a connected graph. So in a physical channel where the amplitude factorizes into two sub-amplitudes $A^{(L)}$ and $A^{(R)}$, if both parts contain external scalars, the internal mode has to be a scalar, otherwise it has to be a gluon.

We show that our formula has the correct factorization properties for YMS. Since the scalars carry a copy of color indices that are not shared by the gluons, the single-trace condition implies that when the amplitude factorizes into two sub-amplitudes $A^{\rm{(L)}}$ and $A^{\rm{(R)}}$, if both parts contain external scalars, the internal particle has to be a scalar, otherwise it must be a gluon.

To be specific, let us assume that $m$ external particles belong to $A^{\rm{(L)}}$. There are altogether three types of configurations of external states: (i) $A^{\rm{(L)}}$ involves only scalars and $A^{\rm{(R)}}$ both scalars and gluons or vice versa, (ii) both $A^{\rm{(L)}}$ and $A^{\rm{(R)}}$ involve scalars and gluons, %(iii) scalars and gluons are fully separated, and (iv) $A^{(L)}$ involves both scalars and gluons and $A^{(R)}$ only gluons or vice versa.
(iii) $A^{\rm{(L)}}$ or $A^{\rm{(R)}}$ involves only gluons.

General method in studying factorization has been provided in \cite{Cachazo:2013hca} and the supplementary materials therein. In all the above cases, the Parke--Taylor-like factor splits into two smaller ones as usual whenever it involves labels from both sides. In addition, it is also not hard to observe that at leading order the Pfaffian $E$ in \eqref{scaYM} factorizes into Pfaffians of two smaller matrices as well, corresponding to $A^{\rm{(L)}}$ and $A^{\rm{(R)}}$ respectively. As a result, one can straightforwardly observe a scalar propagator in case (i) and (ii), i.e.,
\begin{align}
\text{(i)}\quad&A_{q,r}\longrightarrow A^{\rm{(L)}}_{m+1,0}\,\frac{1}{k_I^2}\,A^{\rm{(R)}}_{q-m+1,r},\\
\text{(ii)}\quad&A_{q,r}\longrightarrow A^{\rm{(L)}}_{l+1,m-l}\,\frac{1}{k_I^2}\,A^{\rm{(R)}}_{q-l+1,r-m+l},
\end{align}
where $l$ is some positive integer satisfying $l<q$. The case (iii) is more interesting: for example, when $A^{\rm{(R)}}$ has only gluons, at the leading order the corresponding $\Psi$ matrix has a non-maximal rank and thus its Pfaffian vanishes. So one has to study the behavior of the original Pfaffian at the sub-leading order. Very nicely, after properly breaking up Lorentz products by inserting an additional polarization vector $\epsilon_I$, it turns out that the sub-leading contributions again splits into a Pfaffian and a reduced Pfaffian, the former corresponding to the side involving scalars, while the latter to the side involving gluons only. As a consequence, the internal mode is a gluon, and the amplitude behaves as
%\begin{align}
%\text{(iii)}\quad&A_{q,r}\longrightarrow \sum_{\epsilon_I}A^{(L)}_{q,1}(\epsilon_I)\,\frac{1}{k_I^2}\,A^{(R)}_{0,r+1}(\epsilon_I),\\
%\text{(iv)}\quad&A_{q,r}\longrightarrow \sum_{\epsilon_I}A^{(L)}_{q,m-q+1}(\epsilon_I)\,\frac{1}{k_I^2}\,A^{(R)}_{0,n-m+1}(\epsilon_I).
%\end{align}
\begin{equation}
\text{(iii)}\quad A_{q,r}\longrightarrow \sum_{\epsilon_I}A^{\rm{(L)}}_{q,m-q+1}(\epsilon_I)\,\frac{1}{k_I^2}\,A^{\rm{(R)}}_{0,n-m+1}(\epsilon_I)\,.
\end{equation}
Hence, we see that the results from our formula agree with what Feynman diagrams dictate. %Detailed discussions of the above analysis is presented in \cite{supplementary}.

\subsection{Einstein--Yang--Mills--Scalar Amplitudes}

Now it becomes apparent how to generalize to the formula for Einstein--Yang--Mills--Scalar (EYMS) amplitudes in the single-trace sector. With $q$ scalars, $r$ gluons and $s$ gravitons, the formula can be written as
\be\label{scaEYM}
\mathcal{I}^{\rm{sEYM}}_{q,r,s}=\mathcal{C}_{q{+}r}^{U(N)}\, \mathcal{C}_q^{U(\tilde N)}\, E(\epsilon_{q{+}1},\ldots, \epsilon_n)\,E(\tilde{\epsilon}_{q{+}r{+}1},\ldots,\tilde{\epsilon}_n)\,,
\ee
where for the last $s$ particles, we allow the second copy of polarizations to be different, thus they can be gravitons, B-fields or dilatons. For $q=0$ or $s=0$, it becomes \eqref{EYM} or \eqref{scaYM} respectively, and for $r=0$ it gives the amplitudes for scalars coupled to gravity~\footnote{Since we focus on the single-trace sector, when one type of particles are absent in the external states, they cannot appear as internal states either.}.

Instead of performing more checks on \eqref{scaEYM}, we will see that its validity is guaranteed by the YMS formula,  \eqref{scaYM}, and KLT relations. Recall that in general, KLT relations relate color-ordered amplitudes with scalars and gluons, to amplitudes generally with (double-colored) scalars, (colored) gluons, and gravitons~\cite{Kawai:1985xq}
\be\mathcal{M}_n(\{ S^{\textbf{a},\tilde{\textbf{a}}}, G^{\textbf{a}}, H\})=\sum_{\alpha,\beta\in S_{(n{-}3)!}}\hspace{-2.5ex}A(\alpha\{S_L{=}(S\cup G)^{\textbf{a}},G_L{=}H\})\, \mathcal{S}[\alpha|\beta]\, A(\beta\{S_R{=}S^{\tilde{\textbf{a}}},G_R{=}G\cup H\})\,,\label{KLT}
\ee
where $\mathcal{S}[\alpha|\beta]$ is the KLT bilinear, and $A(\alpha), A(\beta)$ are partial amplitudes with orderings $\alpha,\beta$. $S,G$ and $H$ denote the label sets for scalars, gluons and gravitons, respectively, where the color indices are collectively denoted by $\textbf{a}, \tilde{\textbf{a}}$. On the RHS, $S_L, G_L$ and $S_R, G_R$ denote the label sets for scalars and gluons of both partial amplitudes, respectively, which are determined by $S,G,H$ (note that for scalars, one copy of color indices still remain).

For $S=G=\emptyset$, KLT relations give pure gravity amplitudes in terms of Yang--Mills amplitudes; in~\cite{Cachazo:2013iea} it was proved that those two formulas are exactly related by the KLT relations, based on ``KLT orthogonality" introduced in \cite{Cachazo:2012da} and proved in~\cite{Cachazo:2013gna}. Here we point out that similarly \eqref{scaEYM} is a direct consequence of \eqref{scaYM} and \eqref{KLT}. Recall that KLT orthogonality states that for solutions $\{\sigma^I\},\{\sigma^J\}$ of the scattering equations,
\be
\sum_{\alpha,\beta\in S_{(n{-}3)!}}\hspace{-2.5ex}C_n(\alpha\{\sigma^I\})\, S[\alpha|\beta]\, C_n (\beta\{\sigma^J\})=\delta_{I,J}\,\det{}'\Phi(\{\sigma^I\})^{1/2}\,\det{}'\Phi(\{\sigma^J\})^{1/2}\,,
\ee
where $\det{}'\Phi$ is the Jacobian for doing all integrals in \eqref{generalformula}. Thus the KLT bilinear of partial amplitudes is given by \eqref{generalformula} with the integrand as a product of remaining building blocks
\be
\mathcal{C}^{U(N)}(S_L)\,E(G_L)\,\mathcal{C}^{U(\tilde{N})}(S_R)\,\tilde{E}(G_R)=\mathcal{C}^{U(N)}(S\cup G)\,E(H)\,\mathcal{C}^{U(\tilde{N})}(S)\,\tilde{E}(G\cup H),
\ee
which is exactly the RHS of \eqref{scaEYM}.

\section{Discussions}

The CHY construction of the S-matrix of gravitons, gluons and scalars in arbitrary dimensions based on the scattering equations leads to a very natural question: What are all quantum field theories whose tree-level S-matrices admit such a formulation? One of the obstacles in the construction of formulas for general theories is that the building blocks used originally in the formulation of pure gravity, pure Yang--Mills and pure scalars seemed to be very rigid.

In this work we have shown that the two building blocks are much more flexible than previously expected. Each one admits a generalization. On the one hand, it is possible to use Pfaffians of matrices made from subsets of particles. On the other hand, Parke--Taylor-like factors of subsets of particles are also allowed. Combining these more flexible blocks we have succeeded in constructing the complete single-trace S-matrix of a $U(N)$ Yang--Mills theory minimally coupled to Einstein gravity (EYM). This is, in fact, one example in a family of theories where scalars, gluons and gravitons are mixed. 

This work leads to several natural directions for future research. For example, we have only discussed single trace contributions to amplitudes of gluons and gravitons. When restricted to only external gluons, we have also presented all double-trace amplitudes. It is natural to expect that simple formulas also exist for all multi-trace amplitudes. Another reason to expect the extension of the scattering equation formalism is the recent work on EYM theories with various amounts of supersymmetry where double-copy structures were exposed and exploited for the computation of amplitudes~\cite{Chiodaroli:2014xia}. The reason is that the double-copy construction has been shown to be closely related to KLT orthogonality~\cite{Cachazo:2012da, Cachazo:2013gna} and the CHY representations in \cite{Cachazo:2013iea}.

Another very pressing direction is the introduction of fermions within the scattering equations framework. Already two different approaches have been put forward in the literature. One comes from the ambitwistor string construction as discussed in the introduction~\cite{Mason:2013sva,Adamo:2013tsa}. The other one comes from the work or Bjerrum-Bohr et.~al.~\cite{Bjerrum-Bohr:2014qwa}, where a prescription for converting string theory amplitudes into formulas localized on the scattering equations was given. As a check of the prescription, several four- and five-particle amplitudes were computed. In both approaches, the RNS formulation of strings serves as a guide. A downside of these directions is that the RNS formulation is not the most efficient way of encoding amplitudes involving fermions. Approaches based on pure spinor techniques have been very successful in dealing with fermions due to the fact that they make spacetime supersymmetry manifest~\cite{Berkovits:2000fe}. A concrete possibility is the combination of the work by Mafra, Schlotterer and Stieberger~\cite{Mafra:2011kj,Mafra:2011nv} and Berkovits' infinite tension limit model \cite{Berkovits:2013xba}. 

As mentioned in the introduction, both the ambitwistor string and the Berkovits' approach contain a heterotic version. One of the virtues of heterotic models is that they posses both Yang--Mills and gravity amplitudes in a way that naturally produces Parke--Taylor-like factors when gluons are present. Unfortunately, the gravity sector in both approaches does not seem to lead to Einstein gravity~\cite{Mason:2013sva,Berkovits:2013xba}. We hope that our formulation of Einstein--Yang--Mills theory will inspire a modification of existing world-sheet models so that they would reproduce the amplitudes presented in this paper.

Finally, as discussed in Subsection \ref{explicitchecks}, for the cases of one graviton and any number of gluons, one can rewrite the integrand of our formula, to give the sum of Yang-Mills amplitudes proposed in \cite{Stieberger:2014cea} (see also \cite{Chen:2010ct}). It would be interesting to find a more direct connection between our approach and that coming from the study of the field theory limit of disk relations in string theory.

{\it Acknowledgements:}
This work is supported by Perimeter Institute for Theoretical Physics. Research at Perimeter Institute is supported by the Government of Canada through Industry Canada and by the Province of Ontario through the Ministry of Research \& Innovation.

\bibliographystyle{apsrev4-1}
\bibliography{arbitrary_dimension_prl}

\end{document}